\documentstyle[aps,multicol]{revtex}
\input{epsf}

\begin{document}
\draft
\title{Topological entanglement entropy}
\author{Alexei Kitaev$^{1,2}$ and John Preskill$^1$}
\address{
$^1$ Institute for Quantum Information, California Institute of Technology, Pasadena, CA 91125, USA\\
$^2$ Microsoft Research, One Microsoft Way, Redmond, WA 98052, USA 
}
\maketitle
\begin{abstract}
We formulate a universal characterization of the many-particle quantum entanglement in the ground state of a topologically ordered two-dimensional medium with a mass gap. We consider a disk in the plane, with a smooth boundary of length $L$, large compared to the correlation length. In the ground state, by tracing out all degrees of freedom in the exterior of the disk, we obtain a marginal density operator $\rho$ for the degrees of freedom in the interior. The von Neumann entropy $S(\rho)$ of this density operator, a measure of the entanglement of the interior and exterior variables, has the form $S(\rho)= \alpha L -\gamma + \cdots$, where the ellipsis represents terms that vanish in the limit $L\to\infty$. The coefficient $\alpha$, arising from short wavelength modes localized near the boundary, is nonuniversal and ultraviolet divergent, but $-\gamma$ is a universal additive constant characterizing a global feature of the entanglement in the ground state. Using topological quantum field theory methods, we derive a formula for $\gamma$ in terms of properties of the superselection sectors of the medium.

\end{abstract}
 
\begin{multicols}{2}

In a quantum many-body system at zero temperature, a {\em quantum phase transition} may occur as a parameter varies in the Hamiltonian of the system. The two phases on either side of a quantum critical point may be characterized by different types of {\em quantum order}; the quantum correlations among the microscopic degrees of freedom have qualitatively different properties in the two phases. Yet in some cases, the phases cannot be distinguished by any local order parameter. 

For example, in two spatial dimensions a system with a mass gap can exhibit {\em topological order} \cite{wen}. The quantum entanglement in the ground state of a topologically ordered medium has global properties with remarkable consequences.  For one thing, the quasiparticle excitations of the system ({\em anyons}) exhibit an exotic variant of indistinguishable particle statistics. Furthermore, in the infinite-volume limit the ground-state degeneracy depends on the genus (number of handles) of the closed surface on which the system resides.  

While it is clear that these unusual properties emerge because the ground state is profoundly entangled, up until now no firm connection has been established between topological order and any quantitative measure of entanglement. In this paper we provide such a connection by relating topological order to von Neumann entropy, which quantifies the entanglement of a bipartite pure state. 

Specifically, we consider a disk in the plane, with a smooth boundary of length $L$, large compared to the correlation length. In the ground state, by tracing out all degrees of freedom in the exterior of the disk, we obtain a marginal density operator $\rho$ for the degrees of freedom in the interior. The von Neumann entropy $S(\rho)\equiv -{\rm tr}\rho\log\rho$ of this density operator, a measure of the entanglement of the interior and exterior variables, has the form 
\begin{equation}
S(\rho)= \alpha L -\gamma + \cdots~,
\label{L-entropy}
\end{equation} 
where the ellipsis represents terms that vanish in the limit $L\to\infty$. The coefficient $\alpha$, arising from short wavelength modes localized near the boundary, is nonuniversal and ultraviolet divergent \cite{bombelli}, but $-\gamma$ (where $\gamma$ is nonnegative) is a universal additive constant characterizing a global feature of the entanglement in the ground state. We call $-\gamma$ the {\em topological entanglement entropy}.

This universal quantity reflects topological properties of the entanglement that survive at arbitrarily long distances, and therefore can be studied using an effective field theory that captures the far-infrared behavior of the medium, namely a topological quantum field theory (TQFT) that describes the long-range Aharonov-Bohm interactions of the medium's massive quasiparticle excitations. We find 
\begin{equation}
\label{gamma-formula}
\gamma= \log {\cal D}~,
\end{equation}
where ${\cal D}\ge 1$ is the {\em total quantum dimension} of the medium, given by
\begin{equation}
{\cal D}=\sqrt{\sum_a d_a^2}~;
\end{equation}
here the sum is over all the superselection sectors of the medium, and $d_a$ is the {\em quantum dimension} of a particle with charge $a$.

Any abelian anyon has quantum dimension $d=1$; therefore, for a model of abelian anyons, ${\cal D}^2$ is simply the number of superselection sectors. Thus for a Laughlin state \cite{laughlin} realized in a fractional quantum Hall system with filling factor $\nu=1/q$ where $q$ is an odd integer, we have ${\cal D}= \sqrt{q}$. For the toric code \cite{toric}, which has four sectors, the topological entropy is $\gamma = \log 2$, as has already been noted in \cite{zanardi}. 

However, nonabelian anyons have quantum dimension greater than one. The significance of $d_a$ (which need not be a rational number) is that the dimension $N_{aaa\cdots a}$ of the fusion vector space spanned by all the distinguishable ways in which $n$ anyons of type $a$ can be glued together to yield a trivial total charge grows asymptotically like the $n$th power of $d_a$. For example, in the $SU(2)_k$ Chern-Simons theory, we have
\begin{equation}
{\cal D}^{-1}= \sqrt{\frac{2}{k+2}}\sin\left(\frac{\pi}{k+2} \right)~.
\end{equation}

\begin{figure}[t]
\begin{center}
\leavevmode
\epsfxsize=2.5in
\epsfbox{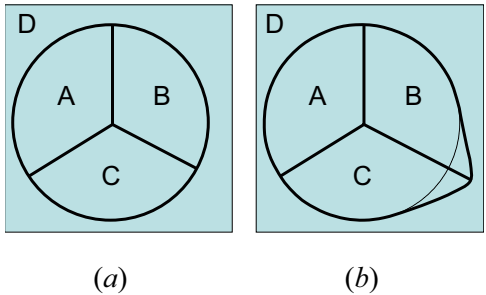}
\end{center}
\caption{($a$) The plane is divided into four regions, labeled $A, B, C, D$, that meet at double and triple intersections. ($b$) Moving the triple intersection where $B,C,D$ meet deforms the regions as shown.}
\label{fig:deform}
\end{figure}

To justify computing the entropy using effective field theory, we require that the boundary have no ``sharp'' features that might be sensitive to short-distance physics; yet for a boundary drawn on a lattice, sharp corners are unavoidable. Furthermore there is an inherent ambiguity in separating the term that scales with the length $L$ from the constant term. We can circumvent these difficulties by the following construction. We now divide the plane into four regions, all large compared to the correlation length, labeled $A,B,C,D$ as in Fig.~\ref{fig:deform}a. Let $S_A$ denote the von Neumann entropy of the density operator $\rho_{A}$ that is obtained from the ground state by tracing out the degrees of freedom outside region $A$, let $S_{AB}$ denote the von Neumann entropy of the density operator $\rho_{AB}$ obtained by tracing out the degrees of freedom outside region $AB\equiv A\cup B$, etc. Then we define the topological entropy $S_{\rm topo}$ as
\begin{equation}
S_{\rm topo}\equiv S_A + S_B +S_C - S_{AB} - S_{BC} - S_{AC} +S_{ABC}~.
\end{equation}
This linear combination of entropies has been strategically chosen to ensure that the dependence on the length of the boundaries of the regions cancels out. For example, the term proportional to the length of the double intersection of $A$ and $D$ appears in $S_A$ and  $S_{ABC}$ with a $+$ sign, and in $-S_{AB}$ and $-S_{AC}$ with a minus sign. Similarly, the double intersection of $A$ and $B$ appears in $S_A$ and $S_{B}$ with a $+$ sign, and in $-S_{AC}$ and in $-S_{BC}$ with a minus sign. (The observation that the ultraviolet divergent terms cancel in a suitably constructed linear combination of entropies has also been exploited in \cite{casini} and applied there to (1+1)-dimensional systems.)

Assuming the behavior eq.~(\ref{L-entropy}) in each term, we find $S_{\rm topo}= -\gamma$. But the advantage of defining $S_{\rm topo}$ using a division into four regions is that we can then argue persuasively that $S_{\rm topo}$ is a topological invariant (dependent only on the topology of how the regions join and not on their geometry) and a universal quantity (unchanged by smooth deformations of the Hamiltonian unless a quantum critical point is encountered). 

To see that $S_{\rm topo}$ is topologically invariant, first consider deforming the boundary between two regions, far from any triple point where three regions meet. Deforming the boundary between $C$ and $D$, say, has no effect on regions $A$, $B$, and $AB$; therefore if all regions are large compared to the correlation length, we expect the changes in $S_A$, $S_B$, and $S_{AB}$ to all be negligible. Thus the change in $S_{\rm topo}$ can be expressed as
\begin{equation}
\Delta S_{\rm topo}= \left(\Delta S_{ABC} - \Delta S_{BC}\right) - \left(\Delta S_{AC}-\Delta S_C\right)~.
\label{topo-change}
\end{equation}
We expect, though, that if the regions are large compared to the correlation length, then appending region $A$ to $BC$  should have a negligible effect on the {\em change} in the entropy, since $A$ is far away from where the deformation is occuring; similarly, appending $A$ to $C$ should not affect the change in the entropy. Thus both terms on the right-hand side of eq.~(\ref{topo-change}) vanish, and $S_{\rm topo}$ is unchanged. The same reasoning applies to the deformation of any other boundary between two regions.

Next consider deforming the position of a triple point, such as the point where $B$, $C$, and $D$ meet as in Fig.~\ref{fig:deform}b. Again we may argue that $S_A$ is unchanged by the deformation. We recall that for a bipartite pure state (like the ground state), the marginal density operators for both subsystems have the same nonzero eigenvalues and therefore the same entropy; thus $S_{ABC}=S_D$ and $S_{BC}=S_{AD}$. We see that the change in $S_{\rm topo}$ can be expressed as
\begin{eqnarray}
\Delta S_{\rm topo}= &&\left(\Delta S_B- \Delta S_{AB}\right) \nonumber\\
+&&\left(\Delta S_C- \Delta S_{AC}\right)\nonumber\\
+&&\left(\Delta S_D- \Delta S_{AD}\right)~.
\label{topo-change-triple}
\end{eqnarray}
All three terms on the right-hand side of eq.~(\ref{topo-change-triple}) vanish because appending $A$ does not affect the change in the entropy. The same reasoning applies when any other triple point moves; we conclude that $S_{\rm topo}$ is unchanged by any deformation of the geometry of the regions that preserves their topology, as long as all regions remain large compared to the correlation length. 

Now, what happens to $S_{\rm topo}$ as the Hamiltonian of the system is smoothly deformed? We assume that the Hamiltonian is a sum of local terms, and that the correlation length remains finite during the deformation (and in fact that the correlation length stays small compared to the size of regions $A,B,C,D$). If the Hamiltonian changes locally in a region far from any boundary, then this change has a negligible effect on the ground state in the vicinity of the boundary, and therefore does not affect $S_{\rm topo}$. If the Hamiltonian changes locally close to a boundary, we can exploit the topological invariance of $S_{\rm topo}$ to first move the boundary far way, then deform the Hamiltonian, and finally return the boundary to its original location. Thus we see that $S_{\rm topo}$ is a universal quantity characteristic of a particular kind of topological order, which remains invariant if no quantum critical point is encountered as the Hamiltonian varies.

\begin{figure}
\begin{center}
\leavevmode
\epsfxsize=3in
\epsfbox{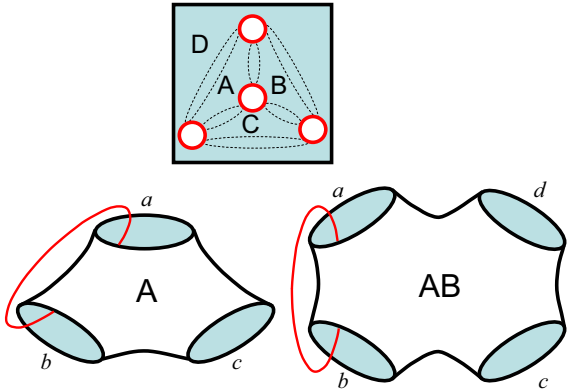}
\end{center}
\caption{The planar medium is glued at spatial infinity to its time-reversal conjugate, and wormholes are attached that connect the two conjugate media at the locations of the triple intersections, creating a sphere with four handles. Each region, together with its image, becomes a sphere with three punctures, and each union of two regions, together with its image, becomes a sphere with four punctures. The punctures carry charges labeled $a,b,c,d$. Anyons that wind around a cycle enclosing a wormhole throat as shown detect a trivial charge.}
\label{fig:pants}
\end{figure}

To facilitate the computation of $S_{\rm topo}$, it is convenient to imagine joining together the planar medium we wish to study with its time-reversal conjugate. We glue together the medium and its conjugate at spatial infinity, and then attach ``wormholes'' that connect the two planes at the positions of the four triple intersections, as indicated in Fig.~\ref{fig:pants}. The resulting closed surface has the topology of a sphere with four handles. If an isolated planar medium is punctured, then massless chiral modes propagate around the edge of the puncture, but the edge states of the medium and its conjugate have opposite chirality, so the edge states acquire masses when the two surfaces are coupled; therefore, the wormholes can be created adiabatically without destroying the mass gap. No anyons are produced during this adiabatic process, so that the mouth of each wormhole carries trivial anyonic charge. 

The boundaries that separate regions in the plane and in its double can be joined through the wormholes as in Fig.~\ref{fig:pants}; then each region of the doubled surface has the topology of a sphere with three punctures, and each union of two adjacent regions becomes a sphere with four punctures. The topological entanglement entropy of the medium and its conjugate are both equal to $S_{\rm topo}$, so that the topological entanglement entropy of the doubled surface is twice $S_{\rm topo}$. The entropy of a region depends only on its topology, so for the doubled surface we have
\begin{equation}
2S_{\rm topo}= 4 S_3 - 3S_4~,
\label{topo-3-4}
\end{equation}
where $S_3$ denotes the entropy for the sphere with three punctures and $S_4$ denotes the entropy for the sphere with four punctures. 

The quantities $S_3$ and $S_4$ can be computed using the appropriate effective field theory, a TQFT \cite{witten}. We use the property that no charge is detected by an anyon that winds around the throat of a wormhole. A cycle that encloses a puncture in the double of (say) region $A$ is complementary to a cycle that winds around the wormhole throat; it follows that the puncture carries charge $a$ with probability
\begin{equation}
p_a = |{\cal S}_1^a|^2= d_a^2/{\cal D}^2~,
\end{equation}
where ${\cal S}_b^a$ is the topological S-matrix of the TQFT, and $1$ denotes the trivial charge. To find the joint probability distribution $p_{abc}$ governing the charges $a,b,c$ on the punctures of the sphere with three punctures, we may use standard TQFT methods to compute the probability $p_{ab\to \bar c}$ that when charges $a$ and $b$ fuse the total charge is $\bar c$. The result is 
\begin{equation}
p_{ab\to \bar c}= N_{abc}d_c /d_a d_b~,
\end{equation}
where $N_{abc}$ is the dimension of the fusion vector space spanned by all the distinguishable ways in which charges $a$, $b$, and $c$ can fuse to yield trivial total charge; it follows that 
\begin{equation}
p_{abc}= p_a p_b \cdot p_{ab\to \bar c}= N_{abc} d_a d_b d_c/{\cal D}^4~.
\end{equation}
Evaluating the entropy in the basis in which each puncture has a definite charge, and summing over all the distinguishable fusion states that occur for specified values of the charges,  we find
\begin{eqnarray}
S_3= &&\sum_{a,b,c} \sum_{\mu=1}^{N_{abc}}-\frac{p_{abc}}{N_{abc}}\log\left(\frac{p_{abc}}{N_{abc}}\right)\nonumber\\
= &&4\log{\cal D} 
- \sum_{a,b,c} p_{abc}\log\left(d_a d_b d_c\right)\nonumber\\
= &&4\log{\cal D} - 3\sum_a p_a\log d_a~.
\end{eqnarray}
For the sphere with four punctures, a similar calculation yields
\begin{equation}
p_{abcd}= p_a p_b p_c \cdot p_{abc\to \bar d}= N_{abcd} d_a d_b d_c d_d/{\cal D}^6
\end{equation}
and 
\begin{eqnarray}
S_4= 6\log{\cal D} - 4\sum_a p_a\log d_a~.
\end{eqnarray}
Plugging into eq.~(\ref{topo-3-4}), we find
\begin{equation}
S_{\rm topo}= 2 S_3 - {3\over 2} S_4 = -\log{\cal D}\equiv -\gamma ~.
\label{topo-formula}
\end{equation}

\begin{figure}
\begin{center}
\leavevmode
\epsfxsize=1in
\epsfbox{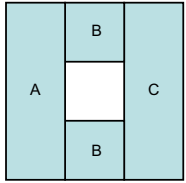}
\end{center}
\caption{If the regions $A,B,C$ have the topology shown, then $S_{\rm topo}= -2\gamma$. }
\label{fig:annulus}
\end{figure}

Eq.~(\ref{topo-formula}) is our main result. Note that it follows if we use eq.~(\ref{L-entropy}) to evaluate the entropy of each region, since $\gamma$ appears four times in the expression for $S_{\rm topo}$ with a negative sign and three times with a positive sign. We also observe that $S_{\rm topo}$ actually depends on the topology of the regions $A,B,C$. For example, consider the arrangement shown in Fig.~\ref{fig:annulus}, in which $B$ and $AC$ both have two connected components, and $ABC$ is not simply connected. Since regions $B$, $AC$, and $ABC$ each have boundaries with two components, now $\gamma$ appears six times with a negative sign and four times with a positive sign, so that $S_{\rm topo}= -2\gamma$.

Using a different approach, we can formulate a simpler but more heuristic derivation of the formula for $\gamma$. First we write the marginal density operator $\rho$ for the disk as $\rho=e^{-\beta H}$. This is just a definition of $H$ and has no other content in itself; furthermore the parameter $T=\beta^{-1}$ is arbitrary --- so we are free to choose it to be small compared to the bulk energy gap of the two-dimensional medium. Now we make a natural but nontrivial assumption: that $H$ can be regarded as the Hamiltonian of a $(1+1)$-dimensional conformal field theory (CFT). This CFT ignores short-distance properties of the bulk medium, and therefore will not account correctly for the term in the entropy proportional to $L$, but it should reproduce correctly the universal constant term. 

To compute the entropy for the case of a disk that contains an anyon with charge $a$ (far from the boundary), we evaluate the partition function $Z_a={\rm tr}_a e^{-\beta H}$ for the associated conformal block of the CFT.  $Z_a$ can be expressed as a path integral on a torus of length $\beta$ in the Euclidean time direction and length $L$ in the spatial direction, in the presence of a Wilson loop carrying anyon charge $a$ that winds through the interior of the torus in the timelike direction. After a modular transformation, we have
\begin{equation}
Z_a= \sum_b {\cal S}_a^b \tilde Z_b,
\end{equation}
where $\tilde Z_b$ is the partition function for the $b$ block on a torus of length $L$ in the Euclidean time direction and length $\beta$ in the spatial direction, and ${\cal S}$ is the modular S-matrix of the CFT, which matches the topological S-matrix of the anyon model. In the limit $L\to\infty$, the sum is dominated by the trivial block $\tilde Z_1$, and we find
\begin{equation}
\log Z_a \approx \log \left({\cal S}_a^1\tilde Z_1\right)\approx \log {\cal S}_a^1 + \frac{\pi}{12}(c+\bar c)L/\beta ~,
\end{equation}
where $c$ and $\bar c$ are the holomorphic and antiholomorphic central charges of the CFT, and $S_a^1=d_a/{\cal D}$ is a topological S-matrix element. Applying the thermodynamic identity $S=-\partial F / \partial T$ (where $F=-T\log Z$ is the free energy),  we then find 
\begin{equation}
S(\rho)=  \frac{\partial}{\partial T}\left( T\log Z\right)=\alpha L - \log \left({\cal D}/d_a\right)~.
\end{equation}
Thus when $a$ is the trivial charge and $d_a=1$, we recover the result of eq.~(\ref{L-entropy}) and (\ref{gamma-formula}). While this derivation is not on so firm a footing as the derivation leading to eq.~(\ref{topo-formula}), it is more transparent and it generalizes readily to the case where the disk contains an anyon. 

We have found an intriguing connection between entanglement entropy and topological order in two dimensions. We note that there are close mathematical ties between the topological entanglement entropy and the (1+1)-dimensional boundary entropy discussed in \cite{affleck}, and we expect that further insights can be derived from studying higher-dimensional analogs of $S_{\rm topo}$. We also hope that our results can provide guidance for the important task of constructing explicit microscopic models that realize topological order.

Results similar to ours have been obtained independently by Levin and Wen \cite{levin}.

We thank Anton Kapustin for discussions.  This work has been supported in part by: the Department of Energy under Grant No. DE-FG03-92-ER40701, the National Science Foundation under Grant No. PHY-0456720, the Army Research Office under Grant Nos. W911NF-04-1-0236 and W911NF-05-1-0294, and the Caltech MURI Center for Quantum Networks under ARO Grant No. DAAD19-00-1-0374.



\end{multicols}
\end{document}